\documentclass[11pt]{article}

\usepackage{amsmath,amssymb,amsfonts,latexsym,amscd,amsthm}
\usepackage{fancyhdr,color,epsfig,setspace,caption,times}
\usepackage[colorlinks=true,plainpages=false,linktocpage,linkcolor=blue,citecolor=green,pagebackref]{hyperref}
\usepackage[letterpaper,left=1.5in,top=1in,bottom=1in,right=1.5in,includeheadfoot,headheight=16.6pt]{geometry}
\usepackage{graphicx}

\usepackage{longtable}
\usepackage{bbm}
\usepackage{dsfont}

\begin{document}

\title{Dominant Strategies in Two Qubit Quantum Computations}
\author{Faisal Shah Khan \\
Khalifa University, Abu Dhabi, UAE}
\maketitle
 
\begin{abstract}
Nash equilibrium is a solution concept in non-strictly competitive, non-cooperative game theory that finds applications in various scientific and engineering disciplines. A non-strictly competitive, non-cooperative game model is presented here for two qubit quantum computations that allows for the characterization of Nash equilibrium in these computations via the inner product of their state space. Nash equilibrium outcomes are optimal under given constraints and therefore offer a game-theoretic measure of constrained optimization of two qubit quantum computations. 
\end{abstract}

\section{Introduction}

The theory of quantum games was originally envisioned by Meyer \cite{Meyer} to be a study of aspects of quantum mechanics such as quantum algorithms via non-cooperative game theory. However, intervening years appear to have blurred the line between the object of study and the tool being used to study it, and a majority of the work in the subject tends to instead study non-cooperative games under a quantum mechanical model. In fact, this pole-reversal if you will, took place early in the development of quantum game theory, starting with the Eisert, Wilkens, and Lewenstein paper \cite{Eisert} and eventually producing a plethora of literature containing results that were really game-theoretic in nature but to which an improper quantum physical significance has been sometimes attached.

Motivated by discussions with colleagues (acknowledged in section \ref{sec:ack} below), Simon J.D. Phoenix and I have recently clarified some of these categorical issues in quantum game theory in \cite{Fkhan1}. We have argued in \cite{Fkhan1} that if the idea behind the merger of quantum physics and game theory is to gain new insights into aspects of quantum physics, then the correct approach to quantum game theory should be ``gaming the quantum'' as envisioned by Meyer and not `` quantizing games'' as proposed by Eisert et al. Let me motivate this argument further here as follows: quantum physicist are familiar and comfortable with quantization of phenomenon, both abstract and those carrying physical significance. For instance, information is quantized by associating the state space of a quantum system (a complex projective Hilbert space) with the set of probability distributions (a convex set). Likewise, space-time is quantized by associating the state space of a quantum system  with a region of space-time (a differentiable manifold). Note that when a phenomenon is quantized, {\it it} is the object of study under a quantum physical model. Therefore, quantum information (more accurately, quantized information) is the study of information processing under a quantum physical model, and quantum field theory is the study of space-time under a quantum physical model. It follows that quantized games study games under a quantum physical model, but not necessarily aspects of quantum physics under a game-theoretic model. Therefore, attaching quantum physical significance to any results in quantized game theory may not always be the most obvious or the most sensible thing to do.

The idea of studying one type of object by analogy with one of another type is an old one; however, it was mathematically formalized in the early part of the twentieth century under the name of category theory where one studies objects in one (mathematical) category via those in another. A fundamentally important feature of this mathematically formal process of forming analogies is a {\it functor}, that is, a map that associates objects in one category to those in another and satisfies certain axioms. John Baez poetically expresses the importance of functors in science in \cite{Baez} as follows: ``...every sufficiently good analogy is yearning to become a functor''. 

What is most relevant to note here about functors is that as with functions, it is important to differentiate between their domain and co-domain. Hence, quantization schemes always map {\it from} the category of Hilbert space {\it into} some other, allowing one to study the objects in the co-domain category via objects in the domain category. It is indeed conceivable that such distinction may sometimes be moot; however, I would argue that if good scientific reasoning has taught us anything it is that one should let such simplification arise naturally from a minimal set of assumptions. 

To make this point further, note that if one insists that a gamed quantum system should collapse to the underlying game upon the introduction of appropriate restrictions \cite{Bleiler}, then one has effectively quantized a game! However, in the absence of such restrictions, one is only gaming the quantum system with the goal of gaining new insights into the quantum system. So quantizing a game is a special case of gaming the quantum, but not necessarily the other way around. 

Category theory has recently become popular among quantum information scientists, as evidenced by works of Abramsky et. al, \cite{Abramsky}, Baez \cite{Baez}, and  Bergholm et al. \cite{Bergholm} that studies quantum informational aspects from a categorical perspective. A similar categorical approach in quantum game theory can only clarify the subject further and offer deeper insights into notions of constrained optimization and equilibrium in quantum systems, for example. 

In this paper, I will build up on my previous work with Simon J.D. Phoenix in \cite{Fkhan1} where we presented a strictly competitive game model for two qubit quantum computations and characterized the corresponding solution concept of mini-max outcomes via the inner product of the Hilbert space of these computations. Strictly competitive games are a special case of the more general class of games known as non-cooperative games. They are so special a case in fact that they are sometimes criticized as being of little practical use in modeling real world phenomena. But history bears witness to the practical usefulness of non-cooperative game theory in areas ranging from Economics to various disciplines of engineering. It can be easily argued that this success of game theory as a practically useful science is due to the work of John Nash in formulating the now ubiquitous solution concept of Nash equilibrium for non-strictly competitive, non-cooperative games. For the sake of further clarity, I point out here that mini-max outcomes are identical to Nash equilibrium in strictly competitive games, but Nash equilibrium need not be a mini-max outcome in general.

Motivated by the significance of Nash equilibrium in game theory and its more traditional applications, I will offer in this paper a non-strictly competitive game model for two qubit quantum computations and establish a notion of Nash equilibrium in such ``quantum games'' using the inner product of the Hilbert space of these computations. It is hoped that this application of the more general non-strictly competitive, non-cooperative game theory to quantum mechanics will produce further new insights in the constrained optimization of quantum mechanisms.”

In the following section \ref{sec:games}, I restate the established notion of functional form of non-cooperative games and the terminology that facilitates the definition of the solution concept of Nash equilibrium. In the same spirit, the notion of dominant strategies is developed in section \ref{sec:domstrat}, followed by notions of quantum games with dominant strategies in section \ref{sec:quantdom} and Nash equilibrium in section \ref{sec:NE} Application to two qubit quantum computations and algorithms is discussed in section \ref{sec:quantcomp}, followed by conclusions and potential future work in section \ref{sec:conc}. 

 \section{Non-cooperative games in normal form}\label{sec:games}
A non-cooperative game in normal form is a function 
\begin{equation}\label{gamma}
\Gamma: \prod X_i \longrightarrow Y  
\end{equation}
with $X_i$ the {\it strategy set} of player $i$ and $Y$ the set of {\it outcomes} with a notion of non-identical preferences of the players defined over these outcomes. In a game with finitely many players, a {\it play} of the game is a tuple of strategies, one per player, $(x_1, x_2, \dots, x_n)$, with $x_i \in X_i$. A play of the game is said to be {\it Nash equilibrium} if unilateral deviation by any player from his choice of strategy will produce an outcome of the game that is less preferable to {\it that} player. In other words, a Nash equilibrium play is one in which each player employs a strategy that is a best reply to those of his opponents. Hence, a Nash equilibrium outcome is an {\it optimal outcome under given constraints}, where the constraints are the non-identical preferences of the players over the outcomes. 

For a more concrete example of a game, consider the following famous game called Prisoner's Dilemma, a two player game with both players having access to two strategies labeled $C$ and $D$. The outcomes of the game are $\left\{o_1, o_2, o_3, o_4 \right\}$ with non-identical preferences of the players over the elements of this set defined as
\begin{equation}\label{prefI}
I: \quad o_3 \succ o_1 \succ o_4 \succ o_2
\end{equation}
\begin{equation}\label{prefII}
II: \quad o_2 \succ o_1 \succ o_4 \succ o_3.
\end{equation}
where $\succ$ represents the notion of ``preferred over''. The game itself can be defined as 
\begin{equation}\label{PD}
P: X_1 \times  X_2  \longrightarrow \left\{o_1, o_2, o_3, o_4 \right\}
\end{equation}
with
\begin{equation}\label{sets}
X_1=X_2=\left\{ C, D \right\}, \quad Y=\left\{o_1, o_2, o_3, o_4 \right\}
\end{equation}
and 
\begin{equation}\label{function}
P(C,C)=o_1, \quad P(C,D)=o_2, \quad P(D,C)=o_3,\quad P(D,D)=o_4
\end{equation}

\section{Two player games with dominant strategies}\label{sec:domstrat}

Games like Prisoner's Dilemma have added structure to the players' preferences. In Prisoner's Dilemma, for player I it is the case that
\begin{equation}\label{dom1}
o_3 \hspace{2mm} {\rm or}  \hspace{2mm} o_4 \succ o_1  \hspace{2mm} {\rm or} \hspace{2mm} o_2
\end{equation}
and for player II it is the case that 
\begin{equation}\label{dom2}
o_2  \hspace{2mm} {\rm or}  \hspace{2mm} o_4 \succ o_1  \hspace{2mm} {\rm or} \hspace{2mm} o_3
\end{equation}
In this situation, the notion of {\it strongly dominant} strategy arises as a strategy that a given player will utilize regardless of what his opponent does. In Prisoner's Dilemma, based on (\ref{function}), player I and II will both always utilize the strategy $D$ regardless of what their opponent does. Hence, appropriate conditions on players' preferences in this case induce players' preferences over their strategic choices, a fact that can be used to compute Nash equilibrium by arguing that a player would never employ a dominated strategy. In Prisoners' Dilemma, this means that the play $(D,D)$ is a Nash equilibrium and the corresponding outcome $o_4$ the Nash equilibrium outcome.

A question of both mathematical and game-theoretic interest could be raised here about the classification of all those two player, two strategy normal form games that entertain dominant strategies (note that the way the function $\Gamma$ maps into $Y$ will influence the existence of dominant strategies) and therefore a Nash equilibrium. Instead of attempting to answer this general question here however, I will consider next the restricted case where the normal form games are quantum physically meaningful.

\section{Two player quantum games with dominant strategies}\label{sec:quantdom}
A {\it quantum game} in normal form is a function $Q$ with $X_i= (H_{d})_i $, a $d$-dimensional Hilbert space representing the strategy set of player $i$, and $Y=H_e$, a $e$-dimensional Hilbert space representing the set of {\it outcomes} with a notion of non-identical preferences of each player defined over its elements. In a game with finitely many players, a {\it play} of the game is a tuple of strategies, one per player, $(x_1, x_2, \dots, x_n)$, with $x_i \in (H_{d})_i$. In functional symbols, 
\begin{equation}\label{quantumgame}
Q: \prod (H_d)_i \longrightarrow H_e
\end{equation}

While it is straight forward to map the functional language of games in normal form to quantum mechanical systems as above, it is not as straight forward to talk of players' preferences over the outcomes, for in any such quantum game we must justify notions of players' preferences within a physical context. This can be achieved by appealing to the concept of observables, that is, elements of an orthogonal basis of $H_e$. For example, consider a two qubit quantum computation $Q$ in the game-theoretic context of Prisoner's Dilemma by setting
\begin{equation}\label{quantident}
(H_d)_1=(H_d)_2=H_2, \quad H_e=H_4.
\end{equation}
The Hilbert space $H_4$ has associated with it four observables in the form of elements of an orthogonal basis, say $B=\left\{ b_1, b_2, b_3, b_4 \right\}$, exactly one of which is the ultimate results of any two qubit quantum computation set up with respect to $B$ and followed by measurement. Given the physical significance of the elements of $B$, one can first define players' preferences over the elements of $B$ as per Prisoner's Dilemma to get
\begin{equation}\label{quantpref1}
{\rm Player \hspace{1mm}  I}: \quad b_3 \succ b_1 \succ b_4 \succ b_2
\end{equation}
\begin{equation}\label{quantpref2}
{\rm Player \hspace{1mm} II}: \quad b_2 \succ b_1 \succ b_4 \succ b_3.
\end{equation}
and further insist that for player I
\begin{equation}\label{quantumdom1}
b_3 \hspace{2mm} {\rm or}  \hspace{2mm}  b_4 \succ b_1  \hspace{2mm} {\rm or} \hspace{2mm} b_2
\end{equation}
and that for player II 
\begin{equation}\label{quantumdom2}
b_2  \hspace{2mm} {\rm or}  \hspace{2mm} b_4 \succ b_1  \hspace{2mm} {\rm or} \hspace{2mm} b_3.
\end{equation}
The inner-product of the $H_4$ allows a more general notion of players' preferences to be set up \cite{Fkhan1} via (\ref{quantumdom1}) and (\ref{quantumdom2}) as follows. Player I will prefer an arbitrary outcome, that is, a quantum superposition $p$ of the elements of $B$ in $H_4$ over another $q$ if $p$ is closer to $b_3$ or $b_4$ than $q$ is, and player II will prefer any arbitrary quantum superposition $r$ over another $s$ if $r$ is closer to $b_2$ or $b_4$ than $s$ is. Denote by $\theta_{(,)}$ the geometric distance between two quantum superpositions induced by the inner-product of Hilbert space; then
\begin{equation}\label{genericpref1}
{\rm Player \hspace{1mm} I}: \quad p \succ q \quad {\rm whenever} \quad  \theta_{(p,b_3)} < \theta_{(q,b_3)} \hspace{2mm} {\rm or} \hspace{2mm} \theta_{(p,b_4)} < \theta_{(q,b_4)}
\end{equation}
\begin{equation}\label{genericpref2}
{\rm Player \hspace{1mm} II}: \quad r \succ s \quad {\rm whenever} \quad  \theta_{(r,b_2)} < \theta_{(s,b_2)} \hspace{2mm} {\rm or} \hspace{2mm} \theta_{(r,b_4)} < \theta_{(s,b_4)}
\end{equation}

Any two qubit quantum computation $Q$ for which the elements of ${\rm Im}(Q) \subseteq H_4$, where ${\rm Im}(Q)$ is the image of $Q$, satisfy (\ref{genericpref1}) and (\ref{genericpref2}) can now be referred to as quantum Prisoner's Dilemma, and one can ask for states of qubits in $H_2$, one per player (players' quantum strategies), that are dominant and correspond to Nash equilibrium. Since quantum Prisoner's Dilemma has infinitely many outcomes corresponding to infinitely many possible strategic choices of the players, it is not possible to identify potential dominant strategies by a direct analysis of the added structure of players’ preferences over the outcomes similar to the one that can be performed in Prisoners Dilemma. Instead, the effect of the added structure of players’ preferences over arbitrary outcomes of quantum Prisoner's Dilemma is first captured by (\ref{genericpref1}) and (\ref{genericpref2}) and these conditions are then used to identify Nash equilibrium; finally, one concludes by virtue of (\ref{genericpref1}) and (\ref{genericpref2}) that all quantum strategies realizing Nash equilibrium are necessarily strongly dominant, as the following discussion shows.

\subsection{Nash equilibrium and dominant strategies}\label{sec:NE}

Nash equilibrium in $Q$ can be identified as follows. Suppose $N \in {\rm Im}(Q)$ is a Nash equilibrium arising from the play $(A^*, B^*)$, that is $Q(A^*,B^*)=N$. If player I unilaterally deviates from strategy $A^*$ to another $A$, then the result $q=Q(A,B^*)$ will be less preferable to him than $N$. Similarly, if player II unilaterally deviates from strategy $B^*$ to another $B$, then the result $s=Q(A^*,B)$ will be less preferable to him than $N$.  It follows that for player I, $N \succ q$ for any $q \in {\rm Im}(Q)$, and that for player 2, $N \succ s$ for any $s \in {\rm Im}(Q)$. Therefore, $N$ is a quantum state that satisfies both  (\ref{genericpref1})  and (\ref{genericpref2}) with $N=p=r$. It follows immediately that all Nash equilibrium quantum strategies are necessarily strongly dominant and the players will never employ any other strategy in quantum Prisoner's Dilemma.

\section{Application to quantum computation}\label{sec:quantcomp}

An immediate significance of gaming two qubit quantum computation lies in the ability to talk meaningfully of constrained optimization of these computations. In \cite{Fkhan2}, Simon J.D. Phoenix and I have optimized two qubit quantum computations under constraints arising from a strictly competitive (also known as zero-sum) game model. Strictly competitive games have the property that one player's win is exactly the other player's loss. As such, strictly competitive constraints offer an interesting potential approach to studying Grover's algorithm which searches for an item from a finite collection. One can view the item being searched for as the winning outcome for one player, and all other outcomes as the winning outcome for the other. A Nash equilibrium in such a quantum game would occur at the so-called mini-max outcome. 

Gaming two qubit quantum computations using the game model of Prisoner's Dilemma here gives the first instance of a proper application of non-strictly competitive game theory to a quantum system, resulting in the characterization of qubit states that produce an optimal outcome under such constraints. Classification of two qubit states and quantum computations that are optimal under the constraints of Prisoner's Dilemma can be achieved by a detailed mechanism design approach similar to the one in \cite{Fkhan2}. The initial set up for such an analysis follows below, with the detailed analysis itself left as an excercise for the reader. 

Start with 
\[
B=\left\{  b_{1}=\left| 0\right\rangle \otimes\left|  0 \right\rangle
,b_{2}=\left| 0 \right\rangle \otimes\left| 1 \right\rangle ,b_{3}=\left| 1
\right\rangle \otimes\left| 0 \right\rangle ,b_{4}=\left| 1 \right\rangle
\otimes\left|  1 \right\rangle \right\} ,
\]
as the standard (orthogonal) computational basis for $H_{4}$
with
\[
\left| 0 \right\rangle = \left(
\begin{array}
[c]{c}%
1\\
0
\end{array}
\right) ; \quad\left|  1 \right\rangle = \left(
\begin{array}
[c]{c}%
0\\
1
\end{array}
\right)
\]
and preferences of the players defined as in expressions (\ref{quantpref1}) - (\ref{quantumdom2}). Consider the two qubit quantum computation
\[
U=\left(
\begin{array}
[c]{cccc}%
U_{11} & U_{12} & U_{13} & U_{14}\\
U_{21} & U_{22} & U_{23} & U_{24}\\
U_{31} & U_{32} & U_{33} & U_{34}\\
U_{41} & U_{42} & U_{43} & U_{44}%
\end{array}
\right)
\]
as a two player, two strategy quantum game represented
with respect to the basis $B$. Also, let $B^{\prime}=\left\{ \left| 0
\right\rangle , \left| 1 \right\rangle \right\} $ be the computational basis
for each $H_{2}$ that constitutes the players' set of strategies. A strategy
for Player I is the choice of a qubit state, say
\[
A=x_{1}\left| 0 \right\rangle +y_{1}\left| 1 \right\rangle
\]
and a strategy for Player II is the choice of qubit state, say
\[
B=x_{2}\left| 0 \right\rangle +y_{2}\left| 1 \right\rangle .
\]

Next, consider a Nash equilibrium play of the quantum game $U$, that is, a play in
which Player I chooses the strategy
\[
A^{*}=\left(
\begin{array}
[c]{c}%
x_{1}^{*}\\
y_{1}^{*}%
\end{array}
\right)
\]
and Player II chooses the strategy
\[
B^{*}=\left(
\begin{array}
[c]{c}%
x_{2}^{*}\\
y_{2}^{*}%
\end{array}
\right) .
\]
The output of the game at this Nash equilibrium is
\begin{equation}
\label{eqn:output}N=U(A^{*},B^{*})=\left(
\begin{array}
[c]{c}%
U_{11}x^{*}_{1}x^{*}_{2}+U_{12}x^{*}_{1}y^{*}_{2}+U_{13}y^{*}_{1}x^{*}%
_{2}+U_{14}y^{*}_{1}y^{*}_{2}\\
U_{21}x^{*}_{1}x^{*}_{2}+U_{22}x^{*}_{1}y^{*}_{2}+U_{23}y^{*}_{1}x^{*}%
_{2}+U_{24}y^{*}_{1}y^{*}_{2}\\
U_{31}x^{*}_{1}x^{*}_{2}+U_{32}x^{*}_{1}y^{*}_{2}+U_{33}y^{*}_{1}x^{*}%
_{2}+U_{34}y^{*}_{1}y^{*}_{2}\\
U_{41}x^{*}_{1}x^{*}_{2}+U_{42}x^{*}_{1}y^{*}_{2}+U_{43}y^{*}_{1}x^{*}%
_{2}+U_{44}y^{*}_{1}y^{*}_{2}%
\end{array}
\right)
\end{equation}
with $\left\|  N \right\| ^{2}=1$ and
\[
\theta_{(N,b_{i})}=\cos^{-1}(\left|  U_{i1}x^{*}_{1}x^{*}_{2}+U_{i2}%
x^{*}_{1}y^{*}_{2}+U_{i3}y^{*}_{1}x^{*}_{2}+U_{i4}y^{*}_{1}y^{*}_{2} \right|
^{2})
\]

\subsection{Nash equilibrium in quantum prisoner's dilemma }\label{sec:NE}

A unilateral deviation on part of Player I from the play $(A^{*}, B^{*})$ to
\textit{any} other $(A,B^{*})$ will produce the output state $\widehat{N}$
such that
\[
\theta_{(\widehat{N},b_{3})} \geq\theta_{(N,b_{3})} \quad  {\rm and} \quad \theta_{(\widehat{N},b_{4})} \geq\theta_{(N,b_{4})}
\]
Because the inverse cosine function is decreasing and the quadratic function is one-to-one and increasing on non-negative inputs, it follows that
\begin{equation}\label{eqn:ineq1}
\left|  U_{31}x_{1}x^{*}_{2}+U_{32}x_{1}y^{*}_{2}+U_{33}%
y_{1}x^{*}_{2}+U_{34}y_{1}y^{*}_{2} \right|  \leq\left|  U_{31}x^{*}_{1}%
x^{*}_{2}+U_{32}x^{*}_{1}y^{*}_{2}+U_{33}y^{*}_{1}x^{*}_{2}+U_{34}y^{*}%
_{1}y^{*}_{2} \right| .
\end{equation}
and that 
\begin{equation}
\label{eqn:ineq2}\left|  U_{41}x_{1}x^{*}_{2}+U_{42}x_{1}y^{*}_{2}+U_{43}%
y_{1}x^{*}_{2}+U_{44}y_{1}y^{*}_{2} \right|  \leq\left|  U_{41}x^{*}_{1}%
x^{*}_{2}+U_{42}x^{*}_{1}y^{*}_{2}+U_{43}y^{*}_{1}x^{*}_{2}+U_{44}y^{*}%
_{1}y^{*}_{2} \right| .
\end{equation}

Likewise, a unilateral deviation on part of Player II from the play $(A^{*},
B^{*})$ to \textit{any} other $(A^{*},B)$ will produce the output state
$\widetilde{N}$ such that
\[
\theta_{(\widetilde{N},b_{2})} \geq\theta_{(N,b_{2})} \quad  {\rm and} \quad \theta_{(\widetilde{N},b_{4})} \geq\theta_{(N,b_{4})}.
\]
Reasoning as in the case of Player I above, the previous two inequalities representing Player II's preferences can be expanded as 
\begin{equation}\label{eqn:ineq3}
\left|  U_{21}x^{*}_{1}x_{2}+U_{22}x^{*}_{1}y_{2}+U_{23}%
y^{*}_{1}x_{2}+U_{24}y^{*}_{1}y_{2} \right|  \leq\left|  U_{21}x^{*}_{1}%
x^{*}_{2}+U_{22}x^{*}_{1}y^{*}_{2}+U_{23}y^{*}_{1}x^{*}_{2}+U_{24}y^{*}%
_{1}y^{*}_{2} \right| .
\end{equation}
and 
\begin{equation}\label{eqn:ineq4}
\left|  U_{41}x^{*}_{1}x_{2}+U_{42}x^{*}_{1}y_{2}+U_{43}%
y^{*}_{1}x_{2}+U_{44}y^{*}_{1}y_{2} \right|  \leq\left|  U_{41}x^{*}_{1}%
x^{*}_{2}+U_{42}x^{*}_{1}y^{*}_{2}+U_{43}y^{*}_{1}x^{*}_{2}+U_{44}y^{*}%
_{1}y^{*}_{2} \right| .
\end{equation}

Focusing back on Player I's preferences and factoring and applying triangle inequality on the right-hand side of
inequality (\ref{eqn:ineq1}) produces 
\[
\left|  U_{31}x_{1}x^{*}_{2}+U_{32}x_{1}y^{*}_{2}+U_{33}y_{1}x^{*}_{2}%
+U_{34}y_{1}y^{*}_{2} \right|  \leq \left|  U_{31}x_{2}^{*}+U_{32}y_{2}%
^{*}\right| \cdot\left|  x_{1}^{*}\right| +\left|  U_{33}x_{2}^{*}+U_{34}%
y_{2}^{*}\right| \cdot\left|  y_{1}^{*}\right|
\]
which can be expressed compactly as
\begin{equation} \label{eqn:shortineq1}
\left|  U_{31}x_{1}x^{*}_{2}+U_{32}x_{1}y^{*}_{2}%
+U_{33}y_{1}x^{*}_{2}+U_{34}y_{1}y^{*}_{2} \right|  \leq P\left|  x_{1}^{*}
\right|  + Q \left|  y_{1}^{*} \right|
\end{equation}
with $P=\left|  U_{31}x_{2}^{*}+U_{32}y_{2}^{*}\right| $ and $Q=\left|
U_{33}x_{2}^{*}+U_{34}y_{2}^{*}\right| $. Applying the same reasoning to
inequality (\ref{eqn:ineq2}) gives
\begin{equation}
\label{eqn:shortineq2}\left|  U_{41}x_{1}x^{*}_{2}+U_{42}x_{1}y^{*}_{2}%
+U_{43}y_{1}x^{*}_{2}+U_{44}y_{1}y^{*}_{2} \right|  \leq P^{\prime}\left|  x_{1}^{*}
\right|  + Q^{\prime} \left|  y_{1}^{*} \right|
\end{equation}
with $P^{\prime}=\left|  U_{41}x_{2}^{*}+U_{42}y_{2}^{*}\right| $ and
$Q^{\prime}=\left|  U_{43}x_{2}^{*}+U_{44}y_{2}^{*}\right| $. Inequalities
(\ref{eqn:shortineq1}) and (\ref{eqn:shortineq2}) can be further simplified by
factoring and applying the triangle inequality to their respective left-hand
sides, giving two cases each:
\begin{equation}
\label{eqn:case1}P|x_{1}| +Q|y_{1}| \leq P| x_{1}^{*}| + Q| y_{1}^{*}|
\end{equation}
or
\begin{equation}
\label{eqn:case2}P|x_{1}| +Q|y_{1}| \geq P| x_{1}^{*}| + Q| y_{1}^{*}|
\end{equation}
and
\begin{equation}
\label{eqn:case3}P^{\prime}|x_{1}| +Q^{\prime}|y_{1}| \leq P^{\prime}|
x_{1}^{*}| + Q^{\prime}| y_{1}^{*}|
\end{equation}
or
\begin{equation}
\label{eqn:case4}P^{\prime}|x_{1}| +Q^{\prime}|y_{1}| \geq P^{\prime}|
x_{1}^{*}| + Q^{\prime}| y_{1}^{*}|.
\end{equation}

Similar reasoning applied to inequalities (\ref{eqn:ineq3}) and (\ref{eqn:ineq4}) for Player II will produce the following collection of inequalities:
\begin{equation}
\label{eqn:case5}S|x_{2}| +T|y_{2}| \leq S| x_{2}^{*}| + T| y_{2}^{*}|
\end{equation}
or
\begin{equation}
\label{eqn:case6} S|x_{2}| + T|y_{2}| \geq S| x_{2}^{*}| + T| y_{2}^{*}|
\end{equation}
and
\begin{equation}
\label{eqn:case7}S^{\prime}|x_{2}| +T^{\prime}|y_{2}| \leq S^{\prime}|
x_{2}^{*}| +T^{\prime}| y_{2}^{*}|
\end{equation}
or
\begin{equation}
\label{eqn:case8}S^{\prime}|x_{2}| +T^{\prime}|y_{2}| \geq S^{\prime}|
x_{2}^{*}| + T^{\prime}| y_{2}^{*}|.
\end{equation}
with 
$S=\left|  U_{21}x_{1}^{*}+U_{23}y_{1}^{*}\right| $, $T=\left|  U_{22}x_{1}^{*}+U_{24}y_{1}^{*}\right| $,  $S^{\prime}=\left|  U_{42}x_{1}^{*}+U_{44}y_{1}^{*}\right| $, and  $T^{\prime}=\left|  U_{42}x_{1}^{*}+U_{44}y_{1}^{*}\right| $.

Inequalities (\ref{eqn:case1}) -  (\ref{eqn:case8}) capture the structure of Nash equilibrium in the class of two qubit quantum computations, $U$, using the notion of dominant strategies from Prisoner's Dilemma. In particular, Each player's dominant, and therefore Nash equilibrium, strategic choice can be analyzed in comparison to {\it all} other strategic choices of the player {\it and} with respect to the parameters of the particular game $U$. That is, one can solve inequalities (\ref{eqn:case1}) -  (\ref{eqn:case8}) for the Nash equilibrium parameters $|x_i^*|$ and $|y_i^*|$ in terms of the paramters $|x_i|$ and $|y_i|$ and the parameters of the game $U$

\section{Conclusions}\label{sec:conc}

Although I gamed two qubit quantum computations here via Prisoner's Dilemma, the same could be done using other non-cooperative two player games such Stag Hunt,  Chicken, Battle of the Sexes, and the full general class of the so-called Hawk-Dove games to which the latter two (and Prisoner's Dilemma) belong. Readers interested in the quantization of these games are refered to \cite{Iqbal, Nawaz, Nawaz1}, respectively.

 Multi-qudit quantum computations could be gamed for constrained optimization using non-cooperative multi-player games. More generally, it is possible to game the quantum to construct more general notions of state distinguishability \cite{Fkhan3} as well as to construct notions of quantum data classification \cite{Fkhan4}. 

\section{Acknowledgment}\label{sec:ack}

I am grateful to Derek Abbott, Steven Bleiler, Azhar Iqbal, and Simon J.D. Phoenix for fruitful discussions. I am also indebted to the referees whose advice has elevated the quality of this paper's presentation and results. 

\singlespace

\end{document}